\documentclass[preprint,aps]{revtex4}

\usepackage{graphicx}
\usepackage{dcolumn}
\usepackage{bm}
\usepackage{latexsym}
\usepackage{amsfonts}
\usepackage{amssymb}
\usepackage{amsmath}
\usepackage[usenames]{color}

\begin{document}
\preprint{KUNS-2192}

\title{Stability of Lovelock Black Holes under Tensor Perturbations}
\author{Tomohiro Takahashi}
\author{Jiro Soda}
\affiliation{Department of Physics,  Kyoto University, Kyoto, 606-8501, Japan
}

\date{\today}

\begin{abstract}
We study the stability of static black holes in the third order Lovelock theory.   
We derive a master equation for tensor perturbations.
 Using the master equation, we analyze the stability
 of Lovelock black holes mainly in seven and eight dimensions. 
 We find there are cases where the linear analysis breaks down.
 If we restrict ourselves to the regime where the linear analysis is
 legitimate, black holes are always stable in seven dimensions.
  However, in eight dimensions, there exists a critical mass
  below which black holes are unstable. 
  Combining our result in the third order Lovelock theory
   with the previous one in Einstein-Gauss-Bonnet theory, 
  we conjecture that small black holes are unstable in any dimensions. 
 The instability found in this paper
 will be important for the analysis of black holes at the LHC.
\end{abstract}

\maketitle

\section{Introduction}

It is believed that string theory is a promising candidate for the theory
 of everything. Remarkably, string theory can be formulated 
 only in ten dimensions. 
 Apparently, it is necessary to reconcile this prediction with our real world
 by compactifying extra-dimensions or by considering braneworld.
Recently, the idea of large extra-dimensions
in the context of the braneworld has been
 advocated~\cite{ArkaniHamed:1998rs,Antoniadis:1998ig,Kokorelis:2002qi}.
Intriguingly, in the presence of the large extra-dimensions,
black holes could be created at the TeV scale~\cite{Giddings:2001bu}.
Hence, the stability of higher dimensional black holes becomes important
since these black holes could be produced  at the Large Hadron Collider (LHC) 
if the spacetime has larger than six dimensions.

The stability of higher dimensional black holes has been 
an active topic since the seminal papers by Kodama  and Ishibashi~\cite{Kodama:2003jz}.
There are at least two directions to be pursued.
One is to study various black holes in Einstein theory.
This direction is necessary because black holes produced at the LHC
are expected to be charged or rotating. 
A numerical study of charged black holes has been done~\cite{Konoplya:2007jv}.
To investigate the stability of rotating black holes,
a group theoretical method is developed~\cite{Murata:2007gv}.
The method is used to study the stability of squashed black
 holes~\cite{Kimura:2007cr,Ishihara:2008re}
 and 5-dimensional rotating black holes~\cite{Murata:2008yx}.
The stability of rotating black holes in more than 5-dimensions is also
studied~\cite{Carter:2005uw,Kunduri:2006qa,Oota:2008uj,Kodama:2008rq}.
The other direction is to consider the stability of black holes
 in more general gravitational theories.
This direction is also important because black holes are produced at 
the Planck scale where Einstein theory would be no longer valid.
In fact, it is known that Einstein theory is merely a low energy limit of
 string theory~\cite{Boulware:1985wk}. In string theory,
  there are higher curvature corrections in addition to
Einstein-Hilbert term~\cite{Boulware:1985wk}. Thus, it 
is natural to extend gravitational theory into those
 with higher power of curvature in higher dimensions.
  It is Lovelock theory 
that belongs to such class of theories~\cite{Lovelock:1971yv,Charmousis:2008kc}.
In Lovelock theory, it is known that
there exist static black hole solutions~\cite{Wheeler:1985nh}. 
Hence, it is natural to suppose black holes produced at the LHC are of this type~\cite{Barrau:2003tk}.
Thus, it is important to study the stability of these Lovelock black holes.

In the case of the second order Lovelock theory, the so-called 
Einstein-Gauss-Bonnet theory, the stability analysis under tensor
perturbations has been performed~\cite{Dotti:2004sh} (see also an earlier work~\cite{Neupane:2003vz}). 
The analysis has been also extended to the scalar and vector 
perturbations by the same group~\cite{Gleiser:2005ra}.
They have shown there exists the scalar mode instability in five dimensions, 
the tensor mode instability in six dimensions, 
and no instability in other dimensions. 
 However, in the dimensions higher than six, we need to incorporate 
 higher order Lovelock terms. Indeed, when we consider black holes at the LHC, 
it is important to consider more higher order Lovelock terms~\cite{Rychkov:2004sf}.
Hence, in this paper, we study the stability of black holes in the third order
Lovelock theory. In this context, our discussion of the stability
of black holes in seven and eight dimensions becomes complete.
In this paper, we will restrict ourselves to the tensor mode analysis
as a first step. Even in this case, we find the instability of small
black holes in eight dimensions.
This suggests the instability of small black holes is generic in Lovelock theory.

The organization of the paper is as follows.
 In Section \ref{seq:2}, we review Lovelock theory and 
introduce Lovelock black hole solutions. 
In Section \ref{seq:3}, we present a master equation for tensor perturbations
in the background of Lovelock black hole.
In Section \ref{seq:4}, we clarify the conditions of applicability
   of the linear analysis and the stability. 
 Then, we discuss the stability of Lovelock black holes for various cases.
 In particular, we prove the stability of black holes under tensor perturbation
 in seven dimensions and find the instability of small black holes 
 in eight dimensions.
 Finally, we summarize our results in Section \ref{seq:5}.    

\section{Lovelock Black Holes}
\label{seq:2}

In this section, we present the third order Lovelock theory.
In the spacetime dimensions $D\leq 6$, the third order Lovelock term
is not relevant. While, for $D\geq 9$, we need to incorporate the
fourth order Lovelock term, which is beyond the scope of this paper. 
Hence, our main objective in this paper is the third order
Lovelock theory in seven and eight dimensions, although
we mention black holes in other dimensions.

In \cite{Lovelock:1971yv}, the most general symmetric, divergence free rank (1,1) tensor 
is constructed out of a metric and its first and second derivatives.
The corresponding Lagrangian is given by
\begin{eqnarray}
  L = \sum_{m=0}^{[(D-1)/2]} c_n {\cal L}_m ,\label{eq:lag} \ ,
\end{eqnarray}
where ${\cal L}_m$ is defined by
\begin{eqnarray}
  {\cal L}_m = \frac{1}{2^m} 
  \delta^{a_1 b_1 \cdots a_m b_m}_{c_1 d_1 \cdots c_m d_m}
  R_{a_1 b_1}{}^{c_1 d_1} \cdots  R_{a_m b_m}{}^{c_m d_m}  \ .
\end{eqnarray}
Here, $[z]$ means the largest integer satisfying 
an inequality $[z] \leq$z,
 $c_n$ is an arbitrary constant and 
 $\delta^{a_1 b_1 \cdots a_m b_m}_{c_1 d_1 \cdots c_m d_m}$ is the 
generalized totally antisymmetric Kronecker delta.

In this paper, we consider up to the third order term in the Lagrangian (\ref{eq:lag}).
From now on, we set $c_0=-2\Lambda$, $c_1=1$, $c_2=\alpha/2$, $c_3=\beta/3$,
 and $c_m=0$ (for $4\leq m$).  
Then, we can get the following equation from this Lagrangian:
\begin{eqnarray}
    0={\cal G}_{a}{}^{b}
 =\Lambda \delta_{a}{}^{b}+G_{a}^{(1)b}+\alpha G_{a}^{(2)b}+\beta G_{a}^{(3)b} \ ,
 \label{basic}
\end{eqnarray}
where $G_{a}^{(1)b}=R_{a}{}^{b}-\frac{1}{2} R\delta_{a}{}^{b}$ is the Einstein tensor.
 The second order Lovelock tensor, i.e., the so-called Gauss-Bonnet tensor,
  $G_{a}^{(2)b}$ is given by
\begin{eqnarray}
   G_{a}^{(2)b}=R_{c a}{}^{d e}R_{d e}{}^{c b}-2R_{d}{}^{c}R_{c a}{}^{d b}
                -2R_{a}{}^{c}R_{c}{}^{b}+RR_{a}{}^{b}
                -\frac{1}{4}\delta_{a}{}^{b}{\cal L}_2 \ .
\end{eqnarray}
The third order Lovelock tensor $G_{a}^{(3)b}$ reads
\begin{eqnarray}
   G_{a}^{(3)b}&=&R^2R_{a}{}^{b}-4R_{a}{}^{b}R_{c}{}^{d}R_{d}{}^{c}+R_{a}{}^{b}R_{c d}{}^{e f}R_{e f}{}^{c d}-4RR_{a c}{}^{b d}R_{d}{}^{c}\nonumber\\
                              &+&8R_{a c}{}^{b d}R_{d e}{}^{c f}R_{f}{}^{e}+8R_{a c}{}^{b d}R_{e}{}^{c} R_{d}{}^{e}-4R_{a c}{}^{b d}R_{e f}{}^{c g}R_{d g}{}^{e f}\nonumber\\
                              &-&4RR_{a}{}^{c}R_{c}{}^{b}+8R_{c}{}^{b}R_{a d}{}^{c e}R_{e}{}^{d}+8R_{a}{}^{c}R_{c}{}^{d}R_{d}{}^{b}-4R_{c}{}^{b}R_{d e}{}^{c f}R_{a f}{}^{d e}\nonumber \\
                              &+&2RR_{a c}{}^{d e}R_{d e}{}^{b c}-4R_{a c}{}^{d e}R_{d e}{}^{b f}R_{f}{}^{c}+4R_{a}{}^{c}R_{d e}{}^{b f}R_{f c}{}^{d e}+2R_{a c}{}^{e f}R_{g h}{}^{b c}R_{e f}{}^{g h}\nonumber\\
                              &-&8R_{a c}{}^{d e}R_{d f}{}^{b c}R_{e}{}^{f}+8R_{a}{}^{c}R_{c d}{}^{b e}R_{e}{}^{d}-8R_{a c}{}^{d e}R_{d f}{}^{b g}R_{g e}{}^{f c}-\frac{1}{6}\delta_{a}{}^{b}{\cal L}_3 \ .
\end{eqnarray}
In the above tensors, the Lagrangians 
\begin{eqnarray}
   {\cal L}_2&=&R^2-4R_{c}{}^{d}R_{d}{}^{c}+R_{c d}{}^{e f}R_{e f}{}^{c d}
\end{eqnarray}
and
\begin{eqnarray}
   {\cal L}_3&=& 
   R^3 -12RR_{c}{}^{d}R_{d}{}^{c}+16R_{c}{}^{d}R_{d}{}^{e}R_{e}{}^{c}
   +2R_{c d}{}^{e f}R_{e f}{}^{g h}R_{g h}{}^{c d}  \nonumber\\
    & &+8R_{c d}{}^{e f}R_{e g}{}^{c h}R_{h f}{}^{d g}+3RR_{c d}{}^{e f}R_{e f}{}^{c d}
      -24R_{c d}{}^{e f}R_{e g}{}^{c d}R_{f}{}^{g}+24R_{c d}{}^{e f}R_{e}{}^{c}R_{f}{}^{d}.
\end{eqnarray}
have appeared. 

As is shown in \cite{Wheeler:1985nh}, 
there exist static black hole solutions. The line element can be expressed as
\begin{eqnarray}
   ds^2=-f(r)dt^2+\frac{1}{f(r)} dr^2+r^2{\bar \gamma}_{i j}dx^idx^j \ ,
   \label{eq:solution}
\end{eqnarray}
where ${\bar \gamma}_{ij}$ is the metric of $n=D-2$-dimensional manifold with
 constant curvature $\kappa=1,0$ or $-1$. The function $f(r)$ is 
 determined by Eq.(\ref{basic}). It is convenient to define a new variable
 $\psi$ as
\begin{eqnarray}
    f(r)=\kappa-r^2\psi(r) \ .
    \label{def}
\end{eqnarray}    
Then, the solution  can be found from the following implicit equation
\begin{eqnarray}
  \frac{\beta}{3}(n-1)(n-2)(n-3)(n-4)\psi^3
  +\frac{\alpha}{2}(n-1)(n-2)\psi^2+\psi-\frac{2\Lambda}{n(n+1)}=\frac{\mu}{r^{n+1}}
         \ ,
\label{eq:poly}
\end{eqnarray}
where $\mu$ is a constant of integration
related to ADM mass~\cite{Myers:1988ze}:
\begin{eqnarray}
	M=\frac{2\mu \pi^{(n+1)/2}}{\Gamma((n+1)/2)} \ ,
      \label{eq:ADM}
\end{eqnarray}
where we used a unit $16\pi G =1$.

In this paper, we will concentrate on the asymptotically flat 
spherically symmetric solutions, i.e.,  $\Lambda=0$ and $\kappa=1$. 
We consider only positive mass black holes  $\mu>0$.
In addition to these, we impose extra assumptions
$\alpha\geq 0$ and $\beta\geq 0$, for simplicity.  

	\begin{figure}
		    \begin{center}
		      \includegraphics[height=6cm, width=8cm]{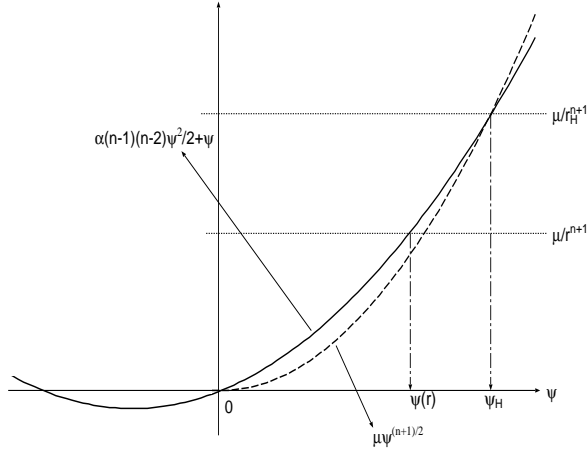}
		      \caption{The intersection between 
                  the solid curve and thin horizontal line determines
                  the solution $\psi = \psi (r)$ for the case $n=4$.
                  Apparently, the infinity $r=\infty$ corresponds to $\psi =0$.
                  The intersection between solid and dashed curve
                  gives a horizon $r_H$.  }
		       \label{fig:1}
		     \end{center}
      \end{figure}
      
\begin{figure}
		  \begin{center}
		    \includegraphics[height=6cm, width=8cm]{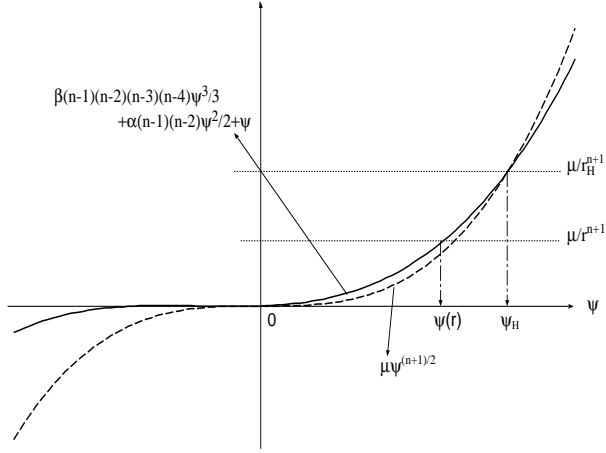}
		    \caption{The same method as $n=4$ case is illustrated
                 for $n=5$ case.}
		   \label{fig:2} 
		  \end{center} 
	\end{figure}

In the case of $n=3$, the theory is reduced to Einstein-Gauss-Bonnet
theory. In this case, we can solve Eq.(\ref{eq:poly}) explicitly
\begin{eqnarray}
\psi = \frac{-1\pm \sqrt{1+\frac{4\alpha \mu}{r^4}}}{2\alpha} \ .
\end{eqnarray}
The upper branch leads 
to the asymptotically flat solution
\begin{eqnarray}
  f(r) = 1 + \frac{r^2}{2\alpha} 
  \left[ 1 - \sqrt{1+\frac{4\alpha\mu}{r^4}}\right]  \ .
\end{eqnarray}
In the case of $n=4$, we have a similar expression.
In Fig.\ref{fig:1}, a graphical method is also explained for this case.
In other dimensions, analytic formula would not be useful.
Hence, we illustrated a graphical method for $n=5$ in Fig.\ref{fig:2}.
We should note that in order to have asymptotically flat solutions
we have to restrict solution of Eq.(\ref{eq:poly}) to the positive one.
 Then, as one can see easily from Eq.(\ref{eq:poly}), 
 $\psi$ goes as $1/r^{n+1}$ in the asymptotic region.
 Hence,  $f(r)$ gives the asymptotically flat metric.

The horizon radius of the asymptotically flat solution is characterized  by $f(r_H)=0$.
 From (\ref{def}), we have a relation $\psi_H=\psi(r_H)=1/r_H^2$.
  Using this relation and (\ref{eq:poly}), we obtain an algebraic equation
\begin{eqnarray}
	\frac{\beta}{3}(n-1)(n-2)(n-3)(n-4)\psi_H^3+\frac{\alpha}{2}(n-1)(n-2)\psi_H^2+\psi_H=\mu \psi_H^{(n+1)/2}.
 \label{eq:rh}
\end{eqnarray}	
This determines $\psi_H$ and hence $r_H$.
 In seven dimensions, one can calculate $\psi_H$ analytically.  
 In fact, substituting $n=5$ into (\ref{eq:rh}), we obtain a simple equation
\begin{eqnarray}
	(8\beta-\mu)\psi_H^2+6\alpha\psi_H+1 = 0 \ . 
\end{eqnarray}
Then, if $\mu\geq 8\beta-9\alpha^2$, the solution is given by
\begin{eqnarray}
	\psi_H=\frac{3\alpha+\sqrt{9\alpha^2-8\beta+\mu}}{\mu-8\beta} \ .
       \label{eq:psih}
\end{eqnarray}
In other dimensions, we do not have analytic formulas.
However, it is easy to obtain solutions of (\ref{eq:rh}) numerically.
 The method to solve Eq.(\ref{eq:rh}) is illustrated
  in Fig.\ref{fig:1} and Fig.\ref{fig:2} in the case of $n=4$ and $n=5$, respectively. 
We note that $\psi$ moves from 0 to $\psi_H$.
 We also notice that the larger $\psi_H$ gives the smaller $\mu$ in this case.
  This is clear from   Fig.\ref{fig:1} and Fig.\ref{fig:2}.
 This is important for later discussion of the stability.      

From the metric (\ref{eq:solution}),
we can calculate the Kretschmann scalar $R_{a b c d}R^{a b c d}$ as
\begin{eqnarray}
	R_{a b c d}R^{a b c d}
      =f^{''2}+2n\frac{f^{'2}}{r^2}+2n(n-1)\frac{(\kappa-f)^2}{r^4} \ ,
\end{eqnarray}
where the prime denotes a derivative with respect to the coordinate $r$. 
Thus, the solution has curvature singularities at $r=0$ or other point
where derivative of $f$ diverges.  
The behavior of a solution can be understood from Eq.(\ref{eq:poly}).
In fact, the solution satisfy the conditions $|1-f|<\infty$, $|f^{'}|<\infty$
 and $|f^{''}|<\infty$ except for $r=0$. Therefore, 
a curvature singularity exists only at $r=0$.

Thus, there is an asymptotically flat solution with a horizon hiding 
a singularity at $r=0$. 
Hence, this solution describes a black hole with the mass $M$ defined in (\ref{eq:ADM}).

\section{master equation for Tensor perturbations}
\label{seq:3}

Understanding of perturbed black holes is important for
the analysis of black holes at the LHC. Of course, the stability of
black holes is a prerequisite for this. 
In the case of higher dimensional black holes,
the stability is not guaranteed in contrast to the 4-dimensional cases. 
To prove the stability, it is convenient to decompose the metric 
under the symmetry of $n$-dimensional symmetric space.
There are scalar, vector, and tensor modes.
In this paper, we will concentrate on the tensor perturbations
as a first step. 

We consider tensor perturbations around the solution (\ref{eq:solution})
\begin{eqnarray}
	\delta g_{a b}=0 \ , \quad \delta g_{a i}=0 \ ,\quad 
      \delta g_{i j}=r^2 \phi(t,r)\bar{h}_{i j}(x^i ) \ , 
\end{eqnarray}
where $a,b=(t,r)$ and $\phi (t,r)$ represents the dynamical degrees of freedom.
Here, $\bar{h}_{ij}$  are defined by
\begin{eqnarray}
	\bar{\nabla}^{k}\bar{\nabla}_{k} \bar{h}_{ij}=\gamma \bar{h}_{ij} \ , \qquad
	\bar{\nabla}^{i} \bar{h}_{ij}=0 \ ,\quad \bar{\gamma}^{ij}\bar{h}_{ij}=0.
\end{eqnarray}
Here, $\bar{\nabla}^{i}$ denotes a covariant derivative with respect to
 $\bar{\gamma}_{ij}$. Here, the eigenvalue is given by 
$\gamma =-\ell (\ell +n-1)+2$, ($\ell =2,3,4 \cdots$) for $\kappa=1$ 
and  negative real number for $\kappa=-1,0$.

As is shown in \cite{Dotti:2004sh}, tensor perturbations around
 the solution (\ref{eq:solution}) in Einstein-Gauss-Bonnet theory can be calculated as
\begin{eqnarray}
  \delta G_{i}^{(1)j}=\left[({\ddot \phi}-f^2\phi^{''})\frac{1}{2f}-\phi^{'}\left(\frac{f^{'}}{2}+\frac{nf}{2r} \right)
          +\frac{\phi}{2r^2}(2\kappa-\gamma)\right]{\bar h}_{i}{}^{j} 
 \label{eq:pert1}
\end{eqnarray}
and
\begin{eqnarray}
  \delta G_{i}^{(2)j} &=& \biggl[({\ddot \phi}-f^2\phi^{''})
  \left(\frac{n-2}{2r^2f}\right)\{-rf^{'}+(n-3)(\kappa-f)\}     \nonumber\\
 && +\phi^{'}\left(\frac{n-2}{2r^3}\right)\{(n-3)[(n-2)(f-\kappa)f-r\kappa f^{'}]
                                          \nonumber\\
    &&  \quad\qquad\qquad\qquad\qquad
        +r^2(f^{'2}+f^{''}f)+(3n-7)rf^{'}f\}\nonumber\\
    &&  +\phi\left(\frac{\gamma-2\kappa}{2r^4}\right)[r^2f^{''}+2(n-3)rf^{'} 
      +(n-3)(n-4)(f-\kappa)]     \biggr]{\bar h}_{i}{}^{j} \ .	
 \label{eq:pert2}
\end{eqnarray}
After a long calculation, we also obtain the third order contribution
\begin{eqnarray}
     \delta G_{i}^{(3)j}&=&\Biggl[\Bigl(\ddot{\phi}-f^2\phi^{''}\Bigr)\frac{1}{2r^4f}(n-2)(n-3)(n-4)(\kappa-f)\left\{(n-5)(\kappa-f)-2rf^{'}\right\}\nonumber\\
		&\ &-\phi^{'}\frac{1}{2r^5}(n-2)(n-3)(n-4)\nonumber\\
		&\ &\hspace{0.5cm}\times\biggl[(f-\kappa)\Bigl\{2r^2ff^{''}+(5n-21)rff^{'}-(n-5)\kappa r f^{'}\nonumber\\
		&\ &\hspace{3cm}+(n-5)(n-4)f(f-\kappa)+4r^2f^{'2}\Bigr\}+2\kappa r^2f^{'2}\biggr] \nonumber\\
	&&+\phi\frac{2\kappa-\gamma}{2r^6}(n-3)(n-4)\nonumber\\
	&&\hspace{0.2cm}\times\Bigl[(f-\kappa)\Bigl\{(n-5)(n-6)(f-\kappa)
      +4(n-5)rf^{'}+2r^2f^{''}\Bigr\}+2r^2f^{'2}\Bigr]\Biggr]\bar{h}_{i}{}^{j} \ . \ \ 
            	 \label{eq:pert3}
\end{eqnarray}
Given the formula (\ref{eq:pert1}), (\ref{eq:pert2}) and (\ref{eq:pert3}), 
the equation can be written as
\begin{equation}
\delta G_{i}^{(1)j}+\alpha\delta G_{i}^{(2)j}+\beta\delta G_{i}^{(3)j}=0 \ .
\end{equation}
Separating the variable $\phi(r,t)=\chi(r) e^{\omega t}$, 
we can deduce the equations to the following form:
\begin{eqnarray}
  - f^2\chi^{''}
  - \left( f^2 \frac{h^{'}}{h}+ \frac{2f^2}{r}
  +  f f^{'} \right) \chi^{'}
   +  \frac{(2\kappa-\gamma)f}{(n-2)r}\frac{h^{'}}{h} \chi = -  \omega^2 \chi \ ,
   \label{eq:pert}
\end{eqnarray}
where
\begin{eqnarray}
	h(r)=r^{n-2}&+&\alpha (n-2)r^{n-4}\{-rf^{'}+(n-3)(\kappa-f)\}\nonumber\\
	               &+&\beta(n-2)(n-3)(n-4)r^{n-6}(\kappa-f)\left\{-2rf^{'}+(n-5)(\kappa-f)\right\} .\label{eq:h}
\end{eqnarray}
The sign of $h$ and its derivative $dh/dr$ is crucial for the stability analysis.

For later purpose, it is useful to express the function $h$ not by the coordinate $r$ 
but by $\psi$ defined in (\ref{eq:poly}). 
From Eq.(\ref{eq:poly}), we can deduce a relation
\begin{eqnarray}
  && \left[\beta(n-4)(n-3)(n-2)(n-1)\psi^2+\alpha(n-2)(n-1)\psi+1 \right] \psi' 
                    \nonumber\\
  && = -\frac{n+1}{r} \left[ \frac{\beta}{3}(n-1)(n-2)(n-3)(n-4)\psi^3
         +\frac{\alpha}{2}(n-1)(n-2)\psi^2+\psi \right] \ .
         \label{prime}
\end{eqnarray}
Using Eq.(\ref{eq:poly}) and the above relation (\ref{prime}),
 we can express (\ref{eq:h}) as a function of $\psi$:
\begin{eqnarray}
	h&=&r^{n-2}\frac{L(\psi)}{6\left[
      \beta(n-4)(n-3)(n-2)(n-1)\psi^2+\alpha(n-2)(n-1)\psi+1 \right]},
      \label{eq:hpsi} 
\end{eqnarray}
where
\begin{eqnarray}
      L(x)&=&2(n-5)(n-4)^2(n-3)^2(n-2)^2(n-1)\beta^2x^4\nonumber\\
          &\ &+4(n-5)(n-4)(n-3)(n-2)^2(n-1)\alpha\beta x^3\nonumber\\
          &\ &+3(n-3)(n-2)\left\{(n-2)(n-1)\alpha^2-8(n-4)\beta\right\}x^2\nonumber\\
          &\ &+6(n-3)(n-2)\alpha x+6  \ . \label{eq:L}
\end{eqnarray}
Similarly, $dh/dr$ is given by
\begin{eqnarray}	
      \frac{dh}{dr}&=&r^{n-3}\frac{K(\psi)}{36\left[
      \beta(n-4)(n-3)(n-2)(n-1)\psi^2+\alpha(n-2)(n-1)\psi+1 \right]^3} \ ,  
      \label{eq:dhpsi}
\end{eqnarray}
where
\begin{eqnarray}
	 K(x)&=&4(n-8)(n-5)(n-4)^4(n-3)^4(n-2)^4(n-1)^3\beta^4x^8\nonumber\\
	     &\ &+16(n-8)(n-5)(n-4)^3(n-3)^3(n-2)^4(n-1)^3\alpha\beta^3x^7\nonumber\\
	     &\ &+2(n-4)^2(n-3)^2(n-2)^3(n-1)^2\bigl\{(n-2)(n-1)(16n^2-163n+469)\alpha^2\nonumber\\
	     &\ &\hspace{3cm}-4(n-4)(n-3)(n+13)(2n-7)\beta\bigr\}\beta^2 x^6\nonumber\\
	     &\ &+6(n-4)(n-3)(n-2)^3(n-1)^2\bigl\{(n-2)(n-1)(5n^2-40n+99)\alpha^2\nonumber\\
	     &\ &\hspace{3cm}+2(n-4)(n-3)(n^2-43n+172)\beta \bigr\}\alpha\beta x^5 \nonumber\\
           &\ &+3(n-3)(n-2)^2(n-1)\bigl\{3(n-5)(n-2)^2(n-1)^2\alpha^4\nonumber\\
           &\ &\hspace{3cm} +2(n-4)(n-2)(n-1)(13n^2-104n+315)\alpha^2\beta\nonumber\\
           &\ &\hspace{4cm}+8(n-4)^2(n-3)(n^2-8n+45)\beta^2 \bigr\}x^4\nonumber\\
           &\ &+12(n-3)(n-2)^2(n-1)\bigl\{3(n-5)(n-2)(n-1)\alpha^2\nonumber\\
           &\ &\hspace{3cm}+2(n-4)(7n^2-21n+80)\beta\bigr\}\alpha x^3\nonumber\\
           &\ &+18(n-2)\bigl\{(n-2)(n-1)(5n^2-25n+42)\alpha^2\nonumber\\
           &\ &\hspace{3cm}+4(n-4)(n-3)(2n^2-n+9)\beta\bigr\}x^2\nonumber\\
           &\ &+108(n-2)(n^2-3n+4)\alpha x+36(n-2)     \ .	
	\label{eq:K}
\end{eqnarray}

In the next section, we will study the stability of black holes
using these formulas (\ref{eq:hpsi}), (\ref{eq:L}), (\ref{eq:dhpsi}),
 and (\ref{eq:K}).

\section{Stability analysis}
\label{seq:4}

Before doing the stability analysis, we will clarify the conditions
for the stability of black holes in Lovelock theory.

First of all, it should be noticed that we have to impose the condition 
\begin{eqnarray}
	h(r)>0 \ ,  \quad ({\rm for}\ r>r_H) \ . \label{eq:assum}
\end{eqnarray}
This is necessary for the linear analysis to be applicable. 
If we choose $f(r)$ to be asymptotically flat solution, one can easily confirm that
 $f(r) \sim \kappa - \mu/r^{n+1}$ for large $r$. 
 It is easy to see $h(r)$ is positive in the asymptotic region.
 Therefore, the condition (\ref{eq:assum}) means that the equation 
$h(r)=0$ has no solution in the region $r > r_H$.
 In the case that there exists $r_0$ such that $h(r_0)=0$ and $r_0>r_H$,
 we encounter a singularity. To see this, let us examine
  (\ref{eq:pert})  around $r_0$. 
 Using approximations $h(r)\sim h' (r_0)(r-r_0)\equiv h'(r_0) y$, 
$f(r)= f(r_0 ) $ and $r=r_0$, we can reduce Eq.(\ref{eq:pert}) into
 the following form:
\begin{eqnarray}
	y \frac{d^2\chi}{dy^2}+\frac{d\chi}{dy}+c \chi=0  \ ,
\end{eqnarray}
where $c$ is a constant. 
The solution near $r_0$ is given by $\chi \sim c_1 + c_2 \log y$, 
where $c_1$ and $c_2$ are constants of integration. 
The solution is singular at $y =0$ for generic perturbations. 
The above calculation means that if the condition (\ref{eq:assum}) is not satisfied,
 then the linear analysis break down at $r_0$. The similar situation
 occurs even in cosmology with higher derivative terms~\cite{Satoh:2007gn}.

When the condition (\ref{eq:assum}) is fulfilled, we can introduce a new variable
\begin{eqnarray}
	\Psi(r)=\chi(r)r\sqrt{h(r)} \ .
\end{eqnarray}
Using $\Psi$ and switching to the coordinate $r^*$, defined by $dr^*/dr=1/f$,
we can rewrite Eq.(\ref{eq:pert}) as
\begin{eqnarray}
	-\frac{d^2\Psi}{dr^{*2}}+V(r(r^*))\Psi=-\omega^2\Psi\equiv E\Psi \ , 
      \label{eq:master}
\end{eqnarray}
where
\begin{eqnarray}
	V(r)=\frac{(2\kappa-\gamma)f}{(n-2)r}\frac{d \ln{h}}{dr}+\frac{1}{r\sqrt{h}}f\frac{d}{dr}\left(f\frac{d}{dr}r\sqrt{h}\right) \label{eq:potential}
\end{eqnarray}
is an effective potential. 

For discussing the stability, the "S-deformation" approach
 is very useful~\cite{Kodama:2003jz, Dotti:2004sh}.  Let us define the operator 
\begin{eqnarray}
	A\equiv -\frac{d^2}{dr^{*2}}+V
\end{eqnarray}
acting on smooth functions defined on $I=(r^{*}_H,\infty)$.
Then, (\ref{eq:master}) is the eigenequation and $E$ is eigenvalue of $A$.
In this case, for any $\varphi$, we can find a smooth function $S$ such that 
\begin{eqnarray}
	(\varphi,A\varphi)=\int_{I} (|D\varphi|^2+\tilde{V}|\varphi|^2)dr^{*},
\end{eqnarray}
where 
\begin{eqnarray}
	D=\frac{d}{dr^{*}}+S  \ , \quad 
	\tilde{V}=V+f\frac{dS}{dr}-S^2  \ .
\end{eqnarray}
Following~\cite{Dotti:2004sh}, we choose $S$ as
\begin{eqnarray}
	S=-f\frac{d}{dr}\ln{(r\sqrt{h})} \ .
\end{eqnarray}
Then, we obtain the formula
\begin{eqnarray}
	(\varphi,A\varphi)
      =\int_{I} |D\varphi|^2dr^{*}+(2\kappa-\gamma)
      \int_{r_H}^{\infty}\frac{|\varphi|^2}{(n-2)r}\frac{d \ln{h}}{dr}dr \ .
       \label{eq:stab}
\end{eqnarray}
Here, the point is that the second term of (\ref{eq:stab}) includes
a factor $2\kappa-\gamma >0$, but $h$ does not include $\gamma$. 
Hence, by taking a sufficiently large $2\kappa -\gamma$, we can 
always make the second term dominant.  

Now, let us show the importance of the sign of $d\ln h /dr$.
If $d\ln h /dr >0$ on $I$, the solution (\ref{eq:solution}) is stable.
This can be understood as follows. 
Note that $2\kappa-\gamma>0$, 
then we have $\tilde{V}>0$ for this case.
 That means $(\varphi,A\varphi)>0$ for arbitrary $\varphi$ if $d \ln{h}/dr>0$ on $I$. 
We choose, for example, $\varphi$ as the lowest eigenstate, then we can conclude that the lowest eigenvalue $E_0$ is positive. 
Thus, we proved the stability. The other way around,
if $d \ln{h} /dr <0$ at some point in $I$, the solution is unstable. 
To prove this, the inequality
\begin{eqnarray}
	\frac{(\varphi,A\varphi)}{(\varphi,\varphi)} \geq E_0 
      \label{eq:ineq}
\end{eqnarray}
is useful. This inequality is correct for arbitrary $\varphi$.
 If $d \ln{h}/dr <0$ at some point in $I$, we can find $\varphi$ such that
\begin{eqnarray}
	 \int_{r_H}^{\infty}\frac{|\varphi|^2}{(n-2)r}\frac{d \ln{h}}{dr}dr<0 \ .
\end{eqnarray}
In this case, (\ref{eq:stab}) is negative for sufficiently large $2\kappa-\gamma$.
 The inequality (\ref{eq:ineq}) implies $E_0<0$ and the solution has unstable modes. 
Thus, we can conclude that the solution is stable if and only if
 $d \ln{h}/dr>0$ on $I$.
 
To summarize, we need to check the sign of $h$ and $dh/dr$
 outside the horizon $r>r_H$ to investigate the stability.
 Note that the condition $h(r) >0$ is necessary for the consistency of the analysis.
To study black holes which do not satisfy this condition, we have to go beyond
the linear analysis.

Now, we are in a position to discuss the stability of Lovelock black holes.
Here, we consider the most important cases, namely,
 asymptotically flat black holes with
  $\alpha\geq 0$, $\beta\geq 0$, $\Lambda=0$, $\mu>0$ and $\kappa=1$.
 Notice that $\psi$ moves on $0\leq \psi \leq \psi_H$ if the solution is asymptotically flat.
 Note that $\psi_H$ is defined in (\ref{eq:rh}). Of course,
 the stability analysis for other cases can be easily done by using our formula.

\subsection{ $\alpha$=0 and $\beta$=0 case} 
	
Let us start with black holes in Einstein theory, i.e., $\alpha=\beta=0$.
For this case, the function $h(r)$ defined by (\ref{eq:h}) becomes $h=r^{n-2}$. 
Apparently,  $h>0$ and $dh/dr>0$ hold. Hence, black holes are stable.
This is consistent with the known result~\cite{Kodama:2003jz}.
	
\subsection{$\alpha>0$ and $\beta$=0 case}

Next, we consider Einstein-Gauss-Bonnet theory~\cite{Dotti:2004sh}.	
From (\ref{eq:hpsi}) and (\ref{eq:L}), $h(r)$ can be read off as
		\begin{eqnarray}
	h=r^{n-2}\frac{(n-3)(n-2)^2(n-1)\alpha^2\psi^2+2(n-3)(n-2)\alpha \psi+2}
      {2\left[ \alpha(n-2)(n-1)\psi+1 \right]}  \ .
		\end{eqnarray}
It is easy to see $h$ is always positive because $\psi\geq 0$ and $\alpha >0$. 
Namely, the linear analysis is always applicable in Einstein-Gauss-Bonnet theory.

Then, we can proceed to check the signature of $dh/dr$
to study the stability of black holes. 
 From (\ref{eq:dhpsi}) and (\ref{eq:K}), we obtain
		\begin{eqnarray}
			\frac{dh}{dr}&=&(n-2)r^{n-3}\frac{F[\psi]}
                  {4\left[ (n-1)(n-2)\alpha\psi+1 \right]^3}  \ , 
		\end{eqnarray}
	where
		\begin{eqnarray}
			F[\psi]=(n-1)^3(n-2)^3(n-3)(n-5)\alpha^4\psi^4+4(n-1)^2(n-2)^2(n-3)(n-5)\alpha^3\psi^3\nonumber\\
	+2(n-1)(n-2)(5n^2-25n+42)\alpha^2\psi^2+12(n^2-3n+4)\alpha\psi+4 \ . \nonumber
		\end{eqnarray}
	Since both $\psi$ and $\alpha$ are positive, 
      $dh/dr$ is always positive if $n\neq 4$. This means
that black holes are stable under tensor purturbations 
other than in six dimensions.
            
In six dimensions, $dh/dr$ becomes
		\begin{eqnarray}
			\frac{dh}{dr}=2r\frac{k(\alpha \psi)}{(6\alpha\psi+1)^3} \ ,
                   \label{eq:38}
		\end{eqnarray}
	where 
		\begin{eqnarray}
			k(x)=-54x^4-36x^3+66x^2+24x+1  \ . 
		\end{eqnarray}
	Because the denominator of (\ref{eq:38}) is always positive,
       the signature of $dh/dr$ is determined by $k(x)$. 
Note that $k(x)$ is always positive in the region
 $0\leq x < \frac{1}{6}(-1+\sqrt{25+10\sqrt{6}})$ and always negative
		in the region $x>\frac{1}{6}(-1+\sqrt{25+10\sqrt{6}})$. 
Since $\psi$ lies in $0\leq \psi \leq \psi_H$, $dh/dr$ is positive 
 when the inequality $\alpha \psi_H < \frac{1}{6}(-1+\sqrt{25+10\sqrt{6}})$ holds. 
 Substituting the marginal value $\alpha \psi_H = \frac{1}{6}(-1+\sqrt{25+10\sqrt{6}})$
 into  Eq.(\ref{eq:rh}), we get the marginal $\mu$:
	\begin{eqnarray}
	\mu_{\rm marginal}=\frac{3\sqrt{6}\left(1+\sqrt{25+10\sqrt{6}}\right)}
      {\left(-1+\sqrt{25+10\sqrt{6}}\right)^{3/2}}\alpha^{3/2} 
			        \simeq  3.98247\alpha^{3/2} \ .
	\end{eqnarray}
It is obvious that black holes are stable if $\mu>3.98247 \alpha^{3/2}$.
		
Summarizing this case, black holes are always stable other than in six dimensions
 and  those with $\mu$ larger than the critical value
  $\mu_{\rm marginal} = 3.98247 \alpha^{3/2}$ are stable in six dimensions.
This is consistent with the previous result~\cite{Dotti:2004sh}.

Thus, we have reproduced known results. This is an evidence
of correctness of our master equation.
Now, we can proceed to obtain new results. 
	
\subsection{$\alpha$=0 and $\beta>0$ case}

Let us consider a special case where the 
Gauss-Bonnet term accidentally vanishes.
For this special case,  we can give rather general results. 
	
Substituting  $\alpha =0$ into (\ref{eq:hpsi}) and (\ref{eq:L}),
we obtain
	\begin{eqnarray}
		h&=&r^{n-2}\frac{l(\sqrt{\beta}\psi)}{\beta(n-4)(n-3)(n-2)(n-1)\psi^2+1}
                   \label{eq:hl}
	\end{eqnarray}
and 
		\begin{eqnarray}
			l(x)=\frac{1}{3}\Bigl\{(n-5)(n-4)^2(n-3)^2(n-2)^2(n-1)x^4
			 -12(n-4)(n-3)(n-2)x^2+3\Bigr\} \ . \ \ \label{eq:l} \ .
		\end{eqnarray}
Similarly, formulas (\ref{eq:dhpsi}) and (\ref{eq:K}) lead to
		\begin{eqnarray}
			 \frac{dh}{dr}&=&(n-2)r^{n-3}\frac{k(\sqrt{\beta}\psi)}
                   {\left[ \beta(n-4)(n-3)(n-2)(n-1)\psi^2+1 \right]^3}
                     \label{eq:dhk}
		\end{eqnarray}
and
	\begin{eqnarray}
		k(x)&=&\frac{1}{9}\Bigl\{(n-8)(n-5)(n-4)^4(n-3)^4(n-2)^3(n-1)^3x^8\nonumber\\
			&\ &\hspace{1cm} -2(n-4)^3(n-3)^3(n-2)^2(n-1)^2(n+13)(2n-7)x^6\nonumber\\
			&\ &\hspace{1cm} +6(n-4)^2(n-3)^2(n-2)(n-1)(n^2-8n+45)x^4\nonumber\\
			&\ &\hspace{1cm} +18(n-4)(n-3)(2n^2-n+9)x^2+9
			            \Bigr\} \ . \label{eq:k}
	\end{eqnarray}
Fortunately, $\beta$ dependence can be absorbed into the scaling factor
in $l$ and $k$. This is the reason why the analysis is relatively simple.      

	\begin{figure}
		    \begin{center}
		      \includegraphics[height=6cm, width=8cm]{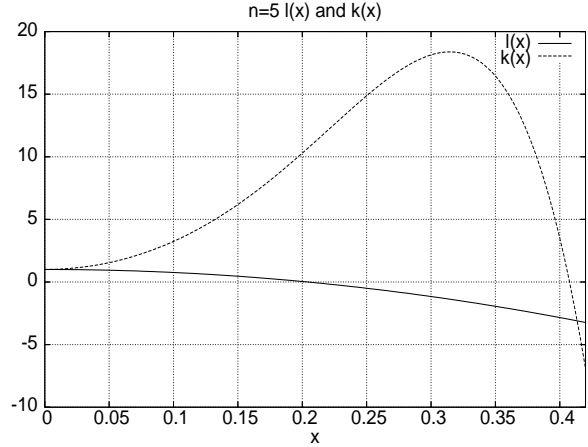}
		      \caption{The behavior of $l(x)$ and $k(x)$ in seven dimensions
                  is shown.
                   Both the lines cross the $x$-axis.
		       Notice that $l(x)$ crosses earlier than $k(x)$. }
		       \label{fig:5}
		     \end{center}
		\end{figure}
            \begin{figure}  
		  \begin{center}
		    \includegraphics[height=6cm, width=8cm]{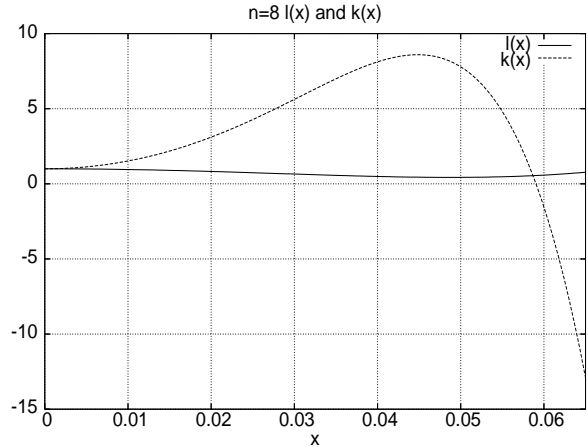}
		    \caption{The behavior of $l(x)$ and $k(x)$ in ten dimensions
                is shown.
                Apparently, $l(x)$ has no positive root. 
                Only $k(x)$ crosses the $x$-axis.}
		   \label{fig:8} 
		  \end{center} 
		\end{figure}

\begin{table}
		\caption{The lowest positive solution of $l(x)=0$ and $k(x)=0$.}
		\begin{center}
		\begin{tabular}{c|c|c}
				n& a & b \\
			\hline
			  	5&$\frac{1}{2\sqrt{6}}\simeq0.2041$ &0.4076 \\
				6&0.1087 &0.1640 \\
				7&0.09129 &0.09129 \\
				8&-& 0.05905 \\
				9&-& 0.04171\\
		\end{tabular}
		\end{center}
		\label{tb:tb1}
\end{table}		
            
To check the sign of $h$ and $dh/dr$,
we need to know the behavior of $l(x)$ and $k(x)$.
These functions for $n=5$ and $n=8$ are plotted in Fig.\ref{fig:5} and Fig.\ref{fig:8},
respectively.
In the case of $n=8$, as is shown in Fig.\ref{fig:8}, $l(x)$ is always positive. 
In fact, we can easily verify $l(x) $ is always positive 
if $n$ is larger than seven by calculating the discriminant of $l(x)=0$.	
Otherwise, the equation $l(x)=0$ has a positive root.

It is convenient to define $a$ as the lowest positive solution of $l(x)=0$ and
 $b$ as that of $k(x)=0$.
 We see that $l(x)$ is positive for $0\leq x < a$ and $k(x)$
 is positive for $0\leq x < b$. 
 The solutions $a$ and $b$ can be calculated numerically 
 and the results are shown in table \ref{tb:tb1}. From this table, we see
$a$ is less than $b$ for $n=5, 6$ and $7$. Note that the linear analysis 
is legitimate in seven, eight and nine dimensions if $\sqrt{\beta}\psi_H < a$. 
Therefore, when the linear analysis is applicable, 
we have the relation $\sqrt{\beta} \psi_H <b$.
Thus, in the cases where the linear analysis is applicable, 
black holes turn out to be stable.
In seven dimensions, for example, comparing (\ref{eq:psih}) and $1/2\sqrt{6}$, we can 
see that the linear analysis is good if $\mu$ is larger than 32$\beta$. 

For $n\geq 8$, $l(x)$ is always positive 
and the solution (\ref{eq:solution}) is stable if $\sqrt{\beta}\psi_H < b$. 
This means that there exists a critical $\mu$ above which
black holes are stable.

To summarize this case, there is a critical $\mu$ in seven, eight and nine dimensions 
due to the limitation of the linear analysis and 
there also exists a critical $\mu$ in $D\geq 10$ due to the instability.

\subsection{$\alpha>0$ and $\beta>0$ case in seven dimensions}

Now we will consider the third order Lovelock theory.
Since the analysis is complicated, we discuss each dimension separately. 		
From now on, we put $\beta=t\alpha^2$ ($t>0$). 
From the perturbative point of view, it is natural to take
 the dimensionless parameter $t$ to be at most order one. 
 However, we keep it arbitrary parameter in our analysis.
            
Now, formulas (\ref{eq:hpsi}) and (\ref{eq:L}) lead to
	\begin{eqnarray}
	     h=r^3\frac{l_t(\alpha \psi)}{24t\alpha^2\psi^2+12\alpha\psi+1} 
	\end{eqnarray}
and
		\begin{eqnarray}
			l_t(x)=12(3-2t)x^2+6x+1    \label{eq:lt} \ .
		\end{eqnarray}
As one can see, $t$ dependence of $l_t (x)$ is a source of complication. 
Apparently, if $t<3/2$, $h$ is positive. 
Similarly,  formulas	(\ref{eq:dhpsi}) and (\ref{eq:K}) become
	\begin{eqnarray}
	   \frac{dh}{dr}=3r^2\frac{k_t(\alpha\psi)}{(24t\alpha^2\psi^2+12\alpha\psi+1)^3} 
	\end{eqnarray}
and
	\begin{eqnarray}
		k_t(x)=6912(3-2t)t^2x^6&+&3456(4-t)tx^5+960t(6+t)x^4\nonumber\\
		             &+&2400tx^3+36(7+6t)x^2+42x+1 \ .  \label{eq:kt}
	\end{eqnarray}
From these expressions, we see the sign of $l_t(r)$ and $k_t(r)$ determines 
that of $h$ and $dh/dr$, respectively.
		
For $0<t\leq\frac{3}{2}$, one can see $l_t(x)$ and $k_t(x)$ are always positive
 because we are considering the case $\alpha\psi\geq 0$ and then $x\geq 0$. 
This means that black holes are stable for any $\mu$. 
For this natural choice of $t$, we do not have any instability under tensor
perturbations. However, if we examine the scalar modes, 
it is likely that the instability shows up. 
           
	\begin{figure}[t]
		    \begin{center}
		      \includegraphics[height=6cm, width=8cm]{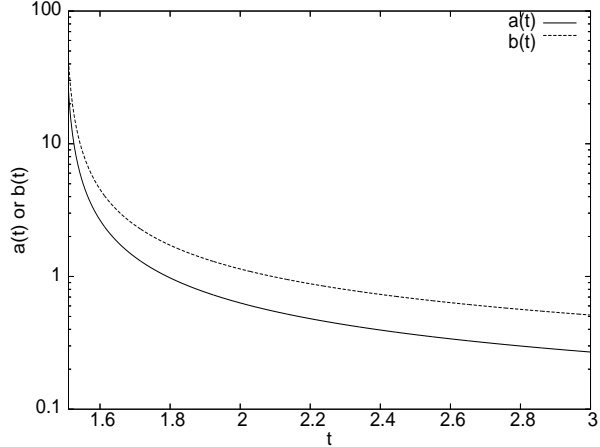}
		      \caption{The graphs of $a(t)$ and $b(t)$ are shown for $t>3/2$ in $D=7$.
                  We clearly see the relation $a(t)< b(t)$. }
		       \label{fig:n5}
		     \end{center}
      \end{figure} 
      		
For $t > \frac{3}{2}$, both $l_t(x)=0$ and $k_t(x)=0$ have positive solutions. 
Let us define $a(t)$ as the lowest solution of $l_t(x)=0$
 and $b(t)$ as that of $k_t(x)=0$. 
We numerically calculated $a(t)$ and $b(t)$ and the results are
 shown in Fig.\ref{fig:n5}. 
It is clear that $l_t(x)$ is positive for $0\leq x < a(t)$ 
and $k_t(x)$ is positive for $0\leq x<b(t)$. 
As can be seen from Fig.\ref{fig:n5}, the relation $a(t)<b(t)$ always holds.
 That means  $k_t(x)$ is positive
  as long as the linear analysis is applicable, i.e., $\alpha\psi_H < a(t)$. 
 Hence, black holes are stable for these cases. 
 Of course, there exist a critical mass for $t>3/2$
  due to the limitation of the linear analysis.

\subsection{$\alpha>0$ and $\beta>0$ case in eight dimensions}

In seven dimensions, we have not seen the instability.
However, in eight dimensions, we will see the instability of small black holes.

 From the formulas (\ref{eq:hpsi}) and (\ref{eq:L}), we obtain
	\begin{eqnarray}
		h=r^4\frac{l_t(\alpha\psi)}{120t\alpha^2\psi^2+20\alpha\psi+1} \ ,
	\end{eqnarray}
where 
	\begin{eqnarray}
			l_t(x)=960t^2x^4+320tx^3+24(5-4t)x^2+12x+1 \ .
	\end{eqnarray}
From other formulas (\ref{eq:dhpsi}) and (\ref{eq:K}), we have
	\begin{eqnarray}
		\frac{dh}{dr}
            =4r^3\frac{k_t(\alpha\psi)}{(120t\alpha^2\psi^2+20\alpha\psi+1)^3} 
            \ , \label{}
	\end{eqnarray}
where
	\begin{eqnarray}
	k_t(x)&=&-2304000t^4x^8-1536000t^3x^7+16000t^2(67-114t)x^6 +24000t(13-10t)x^5 \nonumber\\
		&\ &+240(25+265t+66t^2)x^4+80(206t+15)x^3
			     +180(4+5t)x^2+66x+1  \ .
	\end{eqnarray}
Again, it is clear that the behavior of $l_t(x)$ and $k_t(x)$ 
determines the stability. 

\begin{figure}
		  \begin{center}
		    \includegraphics[height=6cm, width=8cm]{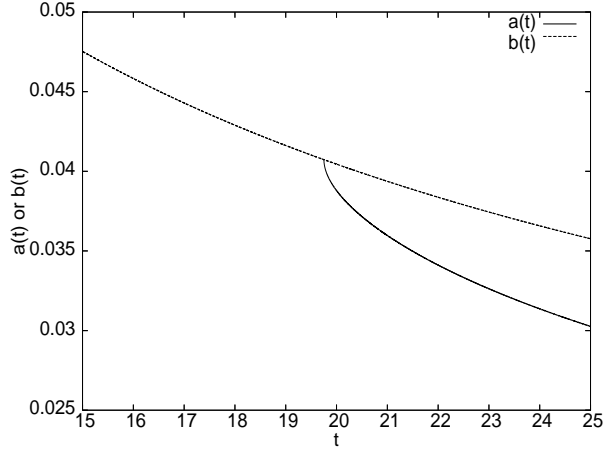}
		    \caption{The behavior of $a(t)$ and $b(t)$ in $D=8$
                is plotted. Below $t=19.752$, $l_t(x)=0$
                has no positive solution. }
		   \label{fig:n6} 
		  \end{center} 
		\end{figure}
		
\begin{figure}
		    \begin{center}
		      \includegraphics[height=6cm, width=8cm]{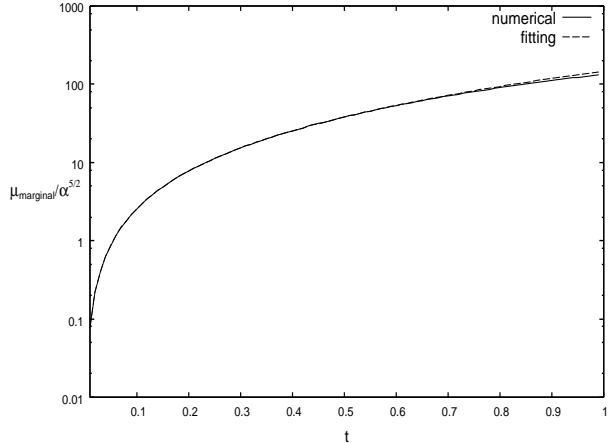}
		      \caption{The marginal value of $\mu_{\rm marginal}$ divided by
                  $\alpha^{5/2}$ as a function of $t$ is plotted 
                  with the solid line.
                  A numerically fit to the data 
                  is plotted with the dashed line.}
		       \label{fig:n6mu}
		     \end{center}
\end{figure}

First, we examine $l_t(x)$. From the discriminant of $l_t(x)=0$, 
we can see that $l_t(x)>0$ in the region $x>0$ when $t$ is less than $19.752$.
 Next, we need to check the sign of $k_t(x)$. 
It is easy to show that $k_t(x)=0$ has positive solutions for arbitrary $t$. 
 We define $b(t)$ as the smallest solution of $k_t(x) =0$. 
 In Fig.\ref{fig:n6}, numerical results for $a(t)$ and $b(t)$ are plotted. 
  From this figure, we can conclude that 
 the solution (\ref{eq:solution}) for $t<19.752$ is stable  if $\alpha\psi_H<b(t)$. 
In Fig.\ref{fig:n6mu}, we numerically calculated the marginal $\mu$,
which satisfies $\alpha \psi_H = b(t)$, as a function of $t$. 
A numerical fit gives us the formula for the marginal $\mu$:
\begin{eqnarray}
  \mu_{\rm marginal} \simeq  (71.4671 \  t^{1.47893}+75.7542 \ t^{2.61059}) \alpha^{5/2} \ .
  \label{main}
\end{eqnarray}
Note that the error of this fitting is about one percent in the range $t<0.5$. 
Therefore, black holes with $\mu$ less than this marginal value (\ref{main})
 is unstable. It should be emphasized that
 the larger the multipole orders $\ell$ becomes, the shorter the time scale
 of the instability becomes. 
This is our main result in this paper. 

We find $l_t(x)=0$ has two positive solutions when $t$ is larger than 19.752. 
We define $a(t)$ as the smaller one.
The solution (\ref{eq:solution}) 
is stable for $t>19.752$ if $\alpha\psi_H<a(t)$. 

To conclude, for $t<19.752$, we have found the instability of small black holes
under tensor perturbations in eight dimensions.
In any case, we have a critical mass below which black holes are unstable
or the linear analysis can not be applicable.

\section{Conclusion}
\label{seq:5}

 We have studied the stability of static black holes in the 
 third order Lovelock theory.   
 We derived a master equation for tensor perturbations.
 Using the master equation, we have reproduced
 known results, which give a check of our master equation.  
 The main purpose of this paper was the stability analysis 
 of Lovelock black holes in seven and eight dimensions. 
 We found there are cases where the linear analysis breaks down.
 If we restrict ourselves to the regime where the linear analysis is
 legitimate, black holes turns out to be always stable in seven dimensions.
 In particular, for a reasonable parameter $t\leq 3/2$, we found
 no critical mass. For $t>3/2$, there is a critical mass due to
 the limitation of the linear analysis. 
 In eight dimensions, for a reasonable range $t<1$,  we found a critical mass
  $\mu_{\rm marginal} \simeq (71.4671 \  t^{1.47893}+75.7542 \ t^{2.61059}) \alpha^{5/2}$
 below which black holes are unstable. 
 Remarkably, the larger the multipole orders $\ell$ becomes,
 the instability  gets the stronger. 
It is interesting to note a similar result is found
 in cosmology in the presence of Gauss-Bonnet term~\cite{Kawai:1998ab}.

 We also examined a special case where the Gauss-Bonnet term 
 disappears accidentally.
 In seven, eight, and nine dimensions, 
 black holes are stable as long as the linear analysis is possible. 
 However, if we regard the breakdown of the linear analysis as an
 indication of the instability, there always exists a critical mass.
 In more than nine dimensions, the linear analysis
 is always legitimate. We have found the instability for small black holes 
 for these dimensions. Thus, in any case, we have a critical mass
 in dimensions $6 \leq D \leq 11$.
 
  The results in seven and eight dimensions
  are similar to that of Einstein-Gauss-Bonnet theory. 
  In Einstein-Gauss-Bonnet theory, there exists the tensor mode 
  instability of small black holes in six dimensions
   and the scalar mode instability in five dimensions.
  Here, we should note that there exists no instability
  in eight dimensions in Einstein-Gauss-Bonnet theory.
  However, in eight dimensions, Einstein-Gauss-Bonnet theory is
  not a complete one in the sense of Lovelock.
  We found the tensor mode instability when we extend the theory to 
  the third order Lovelock theory.
  It is not so absurd to imagine that, if we consider the scalar modes, there exists
  the instability in seven dimensions. 
  If so, it is natural to expect this tendency continues.
  Thus, we conjecture
  that small black holes are unstable in any dimensions
  including ten dimensions. 

 There are many remaining issues to be solved.
First of all, we have to give the analysis of scalar and vector perturbations
to complete our stability analysis.
We also need to understand the meaning of the breakdown of the linear analysis.
Moreover, it is worth investigating the fate of the instability.
As the instability is stronger for higher multipole orders $\ell$,
the resultant geometry would be weird. Finally, it is interesting to consider the
Hawking radiation by taking into account the instability.
The whole picture of the black holes at the LHC would be changed.

\begin{acknowledgements}
JS would like to thank Christos Charmousis for discussion
on Lovelock black holes. 
JS is supported by the Japan-U.K. Research Cooperative Program, 
Grant-in-Aid for  Scientific Research Fund of the Ministry of 
Education, Science and Culture of Japan No.18540262. 
\end{acknowledgements}

\end{document}